\documentclass[amsmath,amssymb,pre,twocolumn,showpacs]{revtex4-1}
\usepackage{amsmath}
\usepackage[applemac]{inputenc}			

\usepackage{graphicx}
\usepackage{amssymb}
\usepackage{bm}
\newcommand{\be}{\begin{equation}} 
\newcommand{\ee}{\end{equation}}
\usepackage{soul}
\usepackage{color}

\begin{document}

\title{Collapse of quasi two dimensional wet granular columns}
\author{Riccardo Artoni}
\altaffiliation{Current address:  L'UNAM, IFSTTAR, Route de Bouaye, CS4, 44344 Bouguenais Cedex, France}
\email{riccardo.artoni@ifsttar.fr}
\affiliation{Dipartimento di Ingegneria Industriale, Universit\`{a} di Padova, Via Marzolo 9, 35131 Padova, Italy}%
\author{Fabio Gabrieli}
\email{fabio.gabrieli@unipd.it}
\affiliation{Dipartimento di Ingegneria Civile Edile e Ambientale, Universit\`{a} di Padova, Via Marzolo 9, 35131 Padova, Italy.}%
\author{Diego Tono}
\affiliation{Dipartimento di Ingegneria Civile Edile e Ambientale, Universit\`{a} di Padova, Via Marzolo 9, 35131 Padova, Italy.}%
\author{Andrea C. Santomaso}
\affiliation{Dipartimento di Ingegneria Industriale, Universit\`{a} di Padova, Via Marzolo 9, 35131 Padova, Italy}%
\author{Simonetta Cola}
\affiliation{Dipartimento di Ingegneria Civile Edile e Ambientale, Universit\`{a} di Padova, Via Marzolo 9, 35131 Padova, Italy.}%

\begin{abstract}
This paper deals with the experimental characterization of the collapse of wet granular columns in the pendular state, with the purpose of collecting data on triggering and jamming phenomena in wet granular media. The final deposit shape and the runout dynamics were studied for samples of glass beads, varying particle diameter, liquid surface tension and liquid amount. 
We show how the runout distance decreases with increasing water amount  (reaching a plateaux for $w> 1\%$) and increases with increasing Bond number, while the top and toe angles, and the final deposit height increase with increasing water amount and decrease with decreasing Bond number.
Dimensional analysis allowed to discuss possible scalings for the runout length and the top and toe angles: a satisfying scaling was found, based on the combination of Bond number and liquid amount. 
\end{abstract}


\maketitle

\section{Introduction}
Triggering and jamming mechanisms are an important research subject for many fields of science, since they contain theoretical and practical implications. In the granular materials research field, jamming and triggering prediction is a key aspect for the control of industrial processes (storage, handling and transport) as well as for the understanding of natural phenomena (landslides, rock avalanches).
Recent research has been carried out on the collapse of granular columns, an unsteady reference problem which involves both triggering and jamming of the flow.  The typical experiment is very simple: An axisymmetric \cite{lube04,lajeunesse04} or rectangular \cite{lube05, siavoshi05b,lajeunesse05,balmforth05} column is allowed to collapse by lifting, respectively, the containing cylinder or a containing wall. The material collapses generally evolving to a pile with a typical inclination slightly lower than the angle of repose. In the cited works, the position of the avalanche which follows the triggering of the motion was usually registered in time to reconstruct collapse dynamics; the final runout distance and heap height were also estimated in order to develop scaling laws for the final deposit. In the various works available in the literature, the effect of the initial aspect ratio, the size and shape of the grains, and the roughness of the bottom surface were evaluated for the case of dry granular materials. For this case, discrete element simulations were also performed with different approaches (classical molecular dynamics or contact dynamics methods) generally confirming the experimental results \cite{zenit05,staron05b}, and evaluating the effect of particle parameters such as interparticle friction and restitution coefficients \cite{staron07}. 
Various models have also been proposed to explain the obtained experimental data \cite{mangeney05,kerswell05,larrieu06,doyle07}. 
All the cited works were concerned with dry cohesionless materials; more recently, Meriaux and Triantafillou \cite{meriaux08} have extended the research to the case of dry cohesive powders, by means of experiments on rectangular column collapse.

In this work we push forward the research on granular column collapse by analyzing the behavior of wet granular materials. Wet granular materials are very interesting because the presence of water in the granular medium greatly affects phenomena even in very small amounts \cite{hornbaker97}. This research represents a first attempt in the investigation of the effect of small quantities of fluid on triggering and jamming.

Different amounts of liquids with different surface tension values were mixed within samples of glass beads of different size ($d_p$ = 2, 3, and 5 mm) focusing on low degrees of saturation (pendular state). Such wet materials were first poured in a rectangular box and then allowed to flow by removing a lateral wall. The dependence of the kinematics and the final state of the system on grain size and water content was particularly investigated.

\section{Materials and Methods}
\subsection{Experimental set-up}
\begin{figure}[t!]
\centering
\includegraphics[width=\columnwidth]{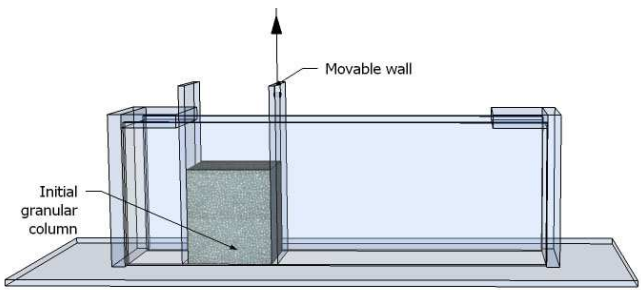}
\caption{\label{telaio} (Color online) Sketch of the experimental channel used for the granular column collapse experiments.}
\end{figure}

Experiments of granular column collapse were performed in a rectangular channel as displayed in Figure \ref{telaio}. The channel was made of transparent glass plates, lateral walls being 35 cm long, 12 cm high, with a gap of 5 cm between them. A 5 x 15 cm glass plate was used as the removable confining wall for the experiment.
As the focus of the present work was to understand the effect of wetting on collapse dynamics, and mainly to investigate different water amounts and particle diameters, the influence of the initial aspect ratio of the column (which was studied, for the dry case, in a number of experimental and numerical works) was not addressed. The initial dimension of the column was therefore kept fixed to length $L_0$=7 cm, height $H_0$=8 cm and thickness $W_0$= 5 cm for all the experiments. The sample was first prepared by slowly adding a certain amount of liquid to the material in a separate vessel; the sample was then manually mixed for a time long enough to ensure homogeneity of liquid bonds. The degree of homogeneity was assessed by means of tomographic measurements, as detailed in the following section. The material was then poured in the experimental box with a funnel. The effective amount of liquid in the material was quantified by weighing the dry, empty preparing container and the container after having poured the spheres into the column. It was important to quantify precisely the amount of water in the sample since the water involved was a very scarce quantity (2 - 17 mL in our tests). The experiment was performed by lifting the moving wall by dropping a sufficiently large weight with a weight-and-pulley system, allowing the column to collapse freely under the action of gravity. Analysing the recorded movies, it was verified that the lifting was perfectly reproducible, with a lifting time of approximately 0.1 seconds.\\
Glass spheres ($\rho_p=$ 2532 $kg/m^3$) in three particle sizes ($d_p=$ 2, 3 and 5 mm) were used for the experiments.
Two different liquids were tested: distilled water and a solution of water with a fluorinated surfactant (1-perfluorotooctyl-3-propan-2-ol) with a concentration of 0.5 g/l, which had the effect of lowering the surface tension of water. Surface tension was measured using a KRUSS K6 tensiometer. Tests with surfactant will be denoted with a `T' in the following.  For each test, 5 repetitions were made. Contact angle measurements of the two fluids on the particles were performed by taking microscopical photos of little drops on a flat glass surface cleaned with ethanol. Measured parameters are reported in Table \ref{table_pars}. It can be noticed that while lowering the surface tension, the surfactant worsens the wetting properties of the solution, inducing an increase of the contact angle.
\begin{table}[h!]
\caption{\label{table_pars} Values of liquid surface tension and  glass-liquid contact angle.}
\begin{center}
\begin{tabular}{cccc}
\hline
&& Distilled water & Water + surfactant  \\
\hline
 $\gamma$ & mN/m &72.75 & 17.10 \\
\hline
$\phi$ & $^o$& 15 & 40 \\
\hline
\end{tabular}
\end{center}
\end{table}
Regarding the amount of liquid, different weights were tested $w=$ 0 (dry), 0.5, 1, 2, 4 \%, expressed in percentages of the solid weight. Such values correspond to a low degree of saturation and belong to the so-called pendular state \cite{newitt58}, where simple liquid bridges between pairs of particles tend to form within the material.

\subsection{Measurement techniques}
\begin{figure}
\centering
\includegraphics[width=\columnwidth]{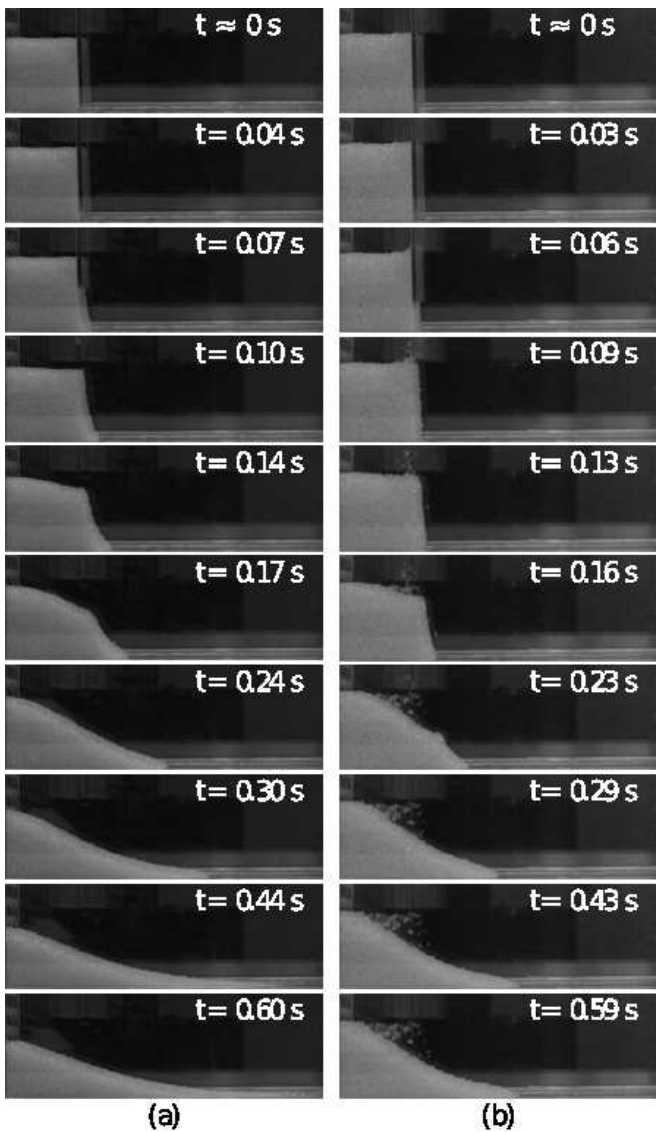}
\caption{\label{cfr}Example image sequences of granular column collapse: (a) $d_p=$ 2 mm, dry case, (b) $d_p=$ 2 mm, $w=$ 1\%, $\gamma=$ 72.75 mN/m (distilled water).}
\end{figure}

During each experiment, the evolution of the system was tracked by taking pictures at fixed time intervals using a CCD camera. In particular, a high resolution semi-professional CCD camera was used (CASIO EX-F1), which was operated at 30 fps, with an exposure time of  1/400 s and a resolution of 2816x2112 pixels. The analysis of the image sequence allowed to track the evolution of the granular mass during time, measuring run-out length, height of the pile, and the angles of the slope at every time-step.
Two examples of the column collapse as recorded in the experiments are reported in Figure \ref{cfr}.  The acquisition and the lifting of the wall were not automatically synchronized in the tests.
In order to analyze the time evolution of the different tests the frame acquisition and the start of the experiments had to be synchronized by determining the zero reference time. As was already stated, an analysis of the moving wall kinematics highlighted that the movement of the wall was fairly reproducible: The friction with the material being nearly negligible, the wall was lifted with a constant acceleration equal to gravity.  For each test the wall position in the first 4-5 frames was fitted to a parabolic equation in order to obtain the zero reference time, which was identified with the time when the wall started to move upwards. The image sequences were then processed in order to obtain descriptive parameters as detailed in the following sections. The frames captured at the very beginning of the collapse allowed to appreciate that the wall lifting was fast enough in order to have no strong effects on the dynamics of the column. Due to the presence of capillary bridges, in wet tests some groups of particles were dragged updwards by the moving wall or remained stick at the lateral walls; this was a phenomenon which could not be avoided but seemed not to have global effects on the collapse.

\begin{figure}[t]
\centering
\includegraphics[width=.8\columnwidth]{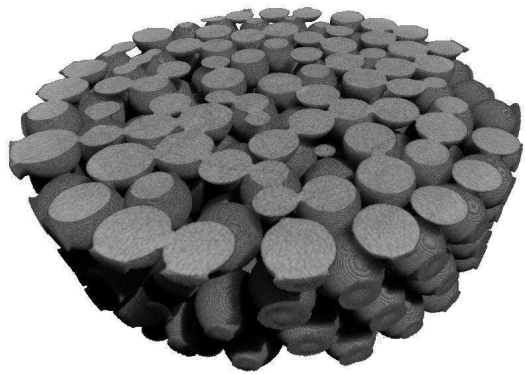}
\caption{\label{tomograf}3D tomographical reconstruction of a sample of glass spheres ($d_p$= 2 mm) with $w=1\%$.}
\end{figure}

Since the behavior of the material depends critically on the number of liquid bridges, it was considered as an important issue to characterize the effect of the preparation procedure on the initial state of the system. In particular  it was needed to gain some knowledge on the degree of uniformity of liquid distribution, and on the number of contacts supporting a liquid bridge. In order to address this problem, some experiments were performed using an  X-ray microtomography scanner (Skyscan 1172). In these tests, first a certain amount of wet material with $w=1\%$ was prepared by manually mixing 200 g of 2 mm glass spheres with a 9 \% solution of KI (potassium iodide) in distilled water. KI was chosen for its X-ray contrasting properties; the surface tension of the solution used was only slightly higher (less than 1 mN/m) than that of pure water. The effect of KI on capillary forces was therefore considered as negligeable. A small portion ($\sim$ 20 g) of the material was then poured in a cylindrical sample holder, which was then introduced in the tomographical apparatus and scanned. Care was taken in order to standardize the mixing procedure; it was however found that a relatively short manual stirring ( $\sim$ 30 s) ensured a proper distribution of the liquid. As a matter of fact, a semiquantitative study was performed on the tomographic images (a reconstruction example is reported in Figure \ref{tomograf}), analyzing three different groups of adjacent slices (located respectively near the top, in the middle, and near the bottom of the sample) in order to evaluate the liquid bridge distribution. For each group of slices (where $\approx $ 80 particles were present), the contact network was analyzed particle by particle, and it was found that only 4\% of the contacts did not support a liquid bridge; many bridges were formed between non contacting particles, and only three triple funicular bridges were found in the whole sample. This analysis was devoted only to understand the efficiency of mixing and allowed to verify that the procedure of manual mixing adopted resulted in a sufficiently homogeneous wet contact network. Further details like the effect of the grain size and liquid properties on the final liquid distribution were not investigated in this work and should be included in further studies.\\

\section{Results}
As was evident from Fig. \ref{cfr}, the presence of a small amount of liquid between the particles strongly affected both the final shape of the deposit and the collapse dynamics. As the typical situations present in the figure show (2 mm glass spheres without liquid and with distilled water at $w=$ 1 \%), the wet material appeared generally to have a longer ``induction'' time, as well as a shorter dynamic phase. Moreover, the deposit had a shorter runout in the wet case, as well as an higher final maximal height. All the tests were quantitatively analyzed by means of descriptive parameters concerning the final deposit shape and runout dynamics as functions of particle size, fluid amount and surface tension, and the results are presented in the following.

\subsection{Effect of particle size, fluid amount and surface tension on the final deposit}
\begin{figure}[ht]
\centering
\includegraphics[width=\columnwidth]{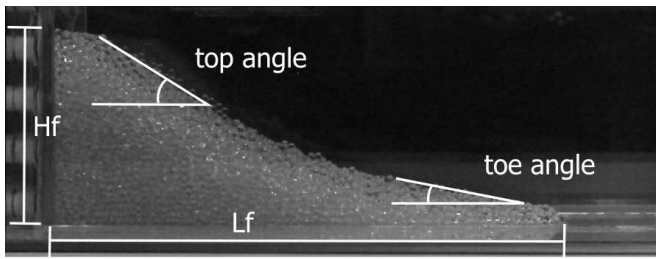}
\caption{\label{sketch-var}  Photograph of a typical final configuration, with the parameters used to characterise the geometrical properties of the deposit.}
\end{figure}

In this section the influence of a wetting fluid on the final configuration attained by the material after run-out is discussed. A typical example of the final deposit shape is given in Fig. \ref{sketch-var}: Notably one can identify a surface slope with two different inclination angles, a higher one near the top ($\theta_{top}$) and a lower one at the end of the deposit (in the following referred to as ``toe angle'', $\theta_{toe}$). The geometry of the final deposit can be also characterised by its vertical and horizontal extent,  namely $H_f$ and $L_f$. In the following the dependence of these parameters on the different variables explored will be discussed. Coherently with the previous literature, lengths are rescaled and non-dimensionalized by studying $H^*=\frac{H_f}{H_0}$ and $L^*=\frac{L_f-L_0}{L_0}$, with $H_0$ and $L_0$ being the initial height and length of the column.

\begin{figure}[ht]
\centering
\includegraphics[width=\columnwidth]{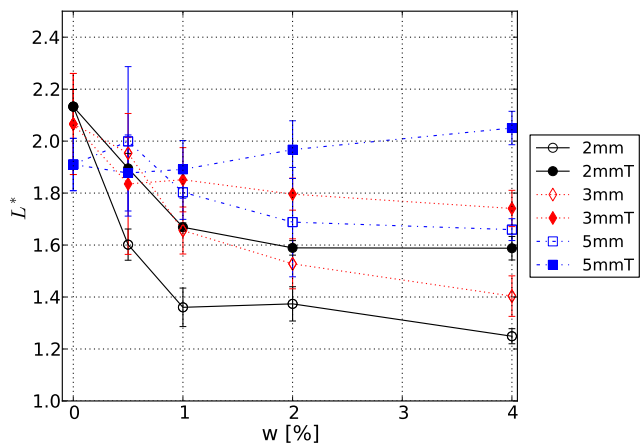}
\caption{\label{Lf}(Color online) Dependence of the rescaled run-out length on liquid content, for different values of particle diameter and liquid surface tension.}
\end{figure}

\begin{figure}[ht]
\centering
\includegraphics[width=\columnwidth]{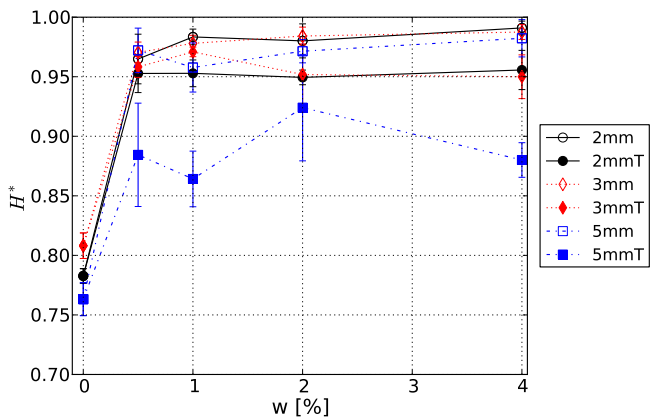}
\caption{\label{Hf}(Color online) Dependence of the rescaled pile height on liquid content, for different values of particle diameter and liquid surface tension.}
\end{figure}

\begin{figure}[ht]
\centering
\includegraphics[width=\columnwidth]{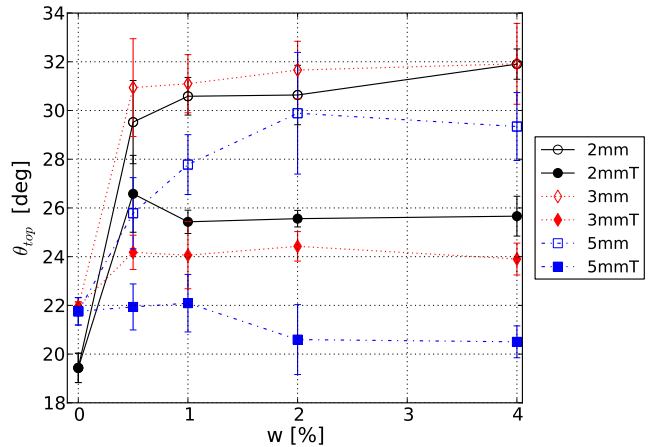}
\caption{\label{top}(Color online) Dependence of the top angle on liquid content, for different values of particle diameter and liquid surface tension.}
\end{figure}

\begin{figure}[ht]
\centering
\includegraphics[width=\columnwidth]{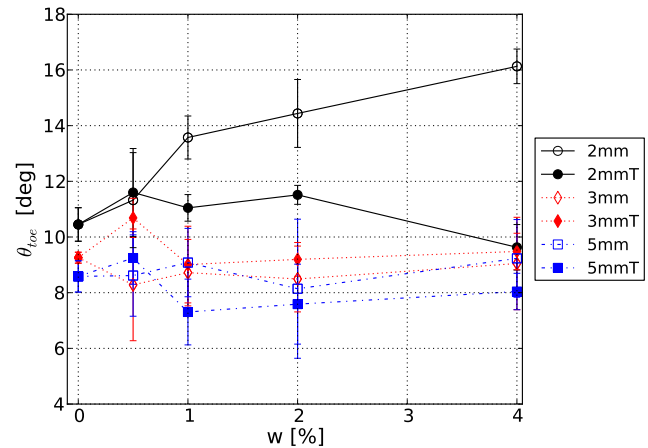}
\caption{\label{toe} (Color online) Dependence of the toe angle on liquid content, for different values of particle diameter and liquid surface tension.}\end{figure}

In previous literature \cite{lube05}, in dry conditions the deposit shape was found not to depend on particle size. The data presented in Fig. \ref{Lf}-\ref{toe} partially confirm this experience: There are few differences in the parameters quantifying the final deposit geometry between the dry tests ($w=0$\%). As  is evident in the following, the presence of capillary forces induced by small amounts of liquids contributes to break the scale invariance, with different behaviours for different particle diameters.

Analyzing the rescaled run-out length, $L^*$, reported in Figure \ref{Lf}, one can notice a separation of behaviours. For all the tests, with the exception of 5 mm particles wetted with the water plus surfactant solution,  $L^*$ decreases when  increasing the amount of fluid, with a plateaux after $w=1\%$; moreover, $L^*$ is longer for low surface tension liquid and for larger particles.
This results can be explained by the fact that the balance between body and capillary forces at particle scale is affected by both particle diameter and surface tension of the liquid. 
Qualitatively (a quantitative discussion will be given in a dedicated section in the following), for larger particles body forces prevail, and the liquid has a lower influence on the statics and dynamics of the granular material;  decreasing surface tension of the liquid has the same effect of increasing particle size.
For the  5 mm particles wetted with water plus surfactant,  $L^*$ is nearly independent of  $w$ below $1 \%$ and has a value equal to the dry case. A slight increase in runout length is found for higher water contents, suggesting that probably lubrication effects come into play.
The presence of a viscous fluid between particles offers a dynamic resistance which, in a dynamic situation, can increase the average interparticle distance, therefore reducing the number of contacts. This makes the capillary forces to be smaller, and reduces the frictional resistance of the material. \emph{Grosso modo,} viscous forces increase for increasing viscosity, particle diameter ($\sim d^2$), and liquid bridge volume ($\sim V^2$ for $V\rightarrow 0$). In its turn, the capillary force is not a simple function of the particle diameter.  For example, following Richefeu et al. \cite{richefeu08}, for equal spheres the capillary force can be approximated by as $F_{cap}=2 \pi \gamma \cos \theta R\; e^{-\frac{\delta}{c (V/R)^0.5}}$, where $R$ is the particle radius, $\delta$ the gap, and $c$ a constant. Given that, depending on system preparation, the gap (and the liquid bridge volume itself) may depend on the particle radius, it is not possible to identify a general behavior. However, for sufficiently dense systems, where liquid bridges are located at the contact points, the force reduces to a linear increasing function of the particle diameter.
The different scalings associated with the interparticle forces are in agreement with the experimental results which show a slight increase in runout length for the case with lowest surface tension and largest particle diameter. In order to evaluate more precisely eventual lubrication effects, other tests should be performed, particularly varying the viscosity \cite{xu07}.

The next shape parameter to be analyzed is the rescaled pile height, $H^*$. As shown in Figure \ref{Hf}, $H^*$ increases as a wetting liquid is added to the material,  with little dependence on liquid amount. For nearly all wet cases the collapse event does not affect the maximum height of the pile, which remains approximately the same as before the event. The only exception is the case with $d_p= 5$ mm and low surface tension solution, which shows intermediate values of $H^*$. 

This saturation behavior  can be explained by thinking in terms of repose angles. 
For simplicity, let's suppose that the material after runout has a triangular shape, with height $H_f$ and length $L_f$. Mass conservation implies that $H_0 L_0=\frac{1}{2}H_f L_f $. It will be obviously that $H_f\leq H_0$. If we suppose that the average slope is approximately equal to the angle of repose, which is a property of the system (depending on the material, the amount of water,  and the type of liquid), we can conclude that there is a limiting value of the angle of repose above which $H_f\approx H_0$.
This limiting angle is $\theta_{crit}=\tan^{-1}(H_0/2L_0)$; in our case this corresponds to $\theta_{crit}=29.7 ^o$.
In Figure \ref{top} one can see how the top angle increases from 20 to 30 degrees when adding a liquid to 2mm spheres, allowing us to suggest that in wet conditions the angle of repose reaches, for the present aspect ratio, the critical value which determines  $H^*\approx 1$. 

Data for the ``top angle'' are displayed in Figure \ref{top}. Here one can clearly see a significant increase of the top angle of the heap when liquid is added to the material, though only a slight dependence is found on the liquid amount. A nice separation of behaviours is found depending on surface tension of the liquid, with the angles belonging to low surface tension tests being generally lower (with again the limit behavior of the $d_p= 5$ mm with surfactant test, which has the lower values with negligible variations with respect to the dry case). This separation was present also in the heap height data.

Regarding the ``toe angle'', $\theta_{toe}$, as displayed in Figure \ref{toe}, some clear information can be extracted from the data: a significant difference between the dry and wet case was not found, apart from the test with the lowest Bond number, that is, the test with $d_p= 2$ mm and water as a liquid. For this case an increasing amount of liquid determines an increasing toe angle. For nearly all the tests the liquid seems not to affect the toe repose angle. This can be related to capillary bridge dynamics: It is at the toe where liquid bridges, if not too strong,  break with higher frequency and therefore it can be reasonable that the heap slope at this point will resemble the dry case.

\subsection{Effect of particle size, fluid amount and surface tension on the run-out dynamics}
\begin{figure}[ht]
\centering
\includegraphics[width=\columnwidth]{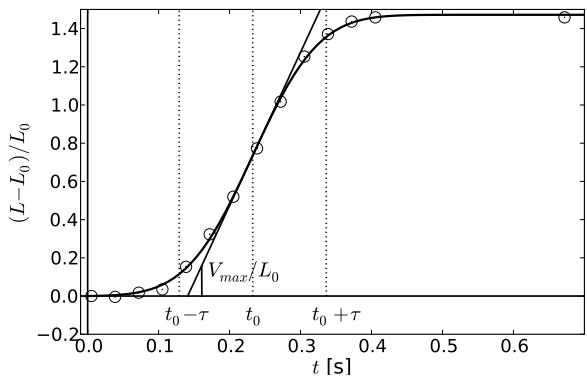}
\caption{\label{dynamic} Example of the runout dynamics for a test with $d_p=$ 2 mm, $w=1. $ \%, pure water as a wetting fluid. The symbols represent the experimental data, while the solid curve is a sigmoid fit on the data.}
\end{figure}

In this second part of the study, we analyze the run-out dynamics of the granular column and its dependence on particle size, amount of wetting fluid and surface tension. As described in the previous literature \cite{lube04,lube05}, when tracking the distance made by the advancing front ($L$) along time, we can identify (i) a region of accelerated motion, (ii) a region of nearly constant velocity, and (iii) a decelerating period at the end of which the flow is stopped. In Refs. \cite{lube04,lube05} it was also shown that for aspect ratios $a=H_0/L_0 < 1.5$ (in this work $a=1.14$), the second region tends to disappear, the dynamics  therefore consisting only of two (accelerating and decelerating) phases.
An example of the dynamics recorded from the experiments is shown in Figure \ref{dynamic}. In fact, for the present data, it was found that the run-out length behavior could be well described by a sigmoidal function with three parameters $V_0$,$t_0$,$\tau$:
\be
L(t)=L_0+\tau V_0 \frac{\sqrt{\pi}}{2}\left[ -\mbox{erf} \left(- \frac{t_0}{\tau}\right) +\mbox{erf} \left(\frac{t-t_0}{\tau} \right)  \right]
\ee
where $t_0$ is the half-time, i.e. the time when $(L-L_0)/(L_f-L_0)=0.5$, $V_0$ is the maximum velocity (occurring at $t=t_0$), $\tau$ is the time span, giving information on the broadness of the runout profile. The derivative of the curve is a gaussian curve centered in $t_0$; the velocity profile in time is therefore nearly symmetrical around $t_0$.
For each set-up, the function was fitted on the data giving estimated values for the parameters, and the 95 \% confidence interval for the parameters was also estimated using Student's t-test.

\begin{figure}[ht]
\centering
\includegraphics[width=\columnwidth]{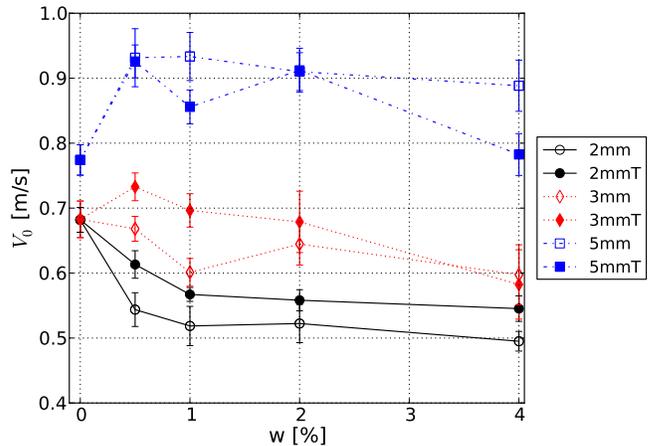}
\caption{\label{Vel}(Color online) Dependence of the maximum velocity attained by the advancing front $V_0$ on liquid content, for different particle size and surface tension values.}
\end{figure}

In Figure \ref{Vel}, values of the maximum attained velocity $V_0$ are reported for all the experimental conditions.
At first it has to be noted that, in dry conditions, again a slight difference exists between tests with different particle diameters. 
As previous literature results suggest particle diameter invariance, this discrepancy may be explained by finite size effects:  Given that the experimental set-up is not scaled with particle diameter, for coarser particles the ratio $W_0/d_p$ probably reaches a critical value such that sidewalls influence differently the dynamics. This has to be kept in mind also in the wet data analysis. When adding liquid, we see a somewhat complex behavior: for $d_p=5$ mm particles, $V_0$ is slightly increased with respect to the dry case, while it decreases for smaller particles. For $d_p=2$ and $3$ mm, velocities in low surface tension tests are slightly larger than the respective high surface tension ones. 

\begin{figure}[ht]
\centering
\includegraphics[width=\columnwidth]{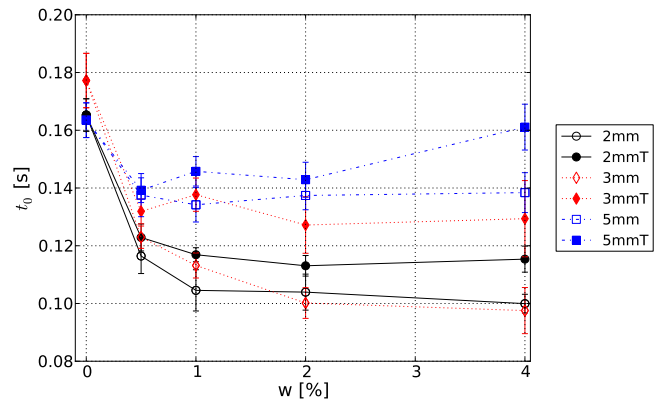}
\caption{\label{t0}(Color online) Dependence of the half-time $t_0$  on liquid content, for different particle size and surface tension values.}
\end{figure}

Fitted values of runout half-time $t_0$ are displayed in Figure \ref{t0}. At first, it is evident that adding liquid $t_0$ decreases, reaching a plateau for $w>1$ \%. The collapse event therefore reaches its half-life more rapidly in wet conditions than in dry ones. This is clearly related to the smaller runout length observed for wet conditions.  Generally, low surface tension tests have a longer half-time than their corresponding high surface tension ones.

\begin{figure}[ht]
\centering
\includegraphics[width=\columnwidth]{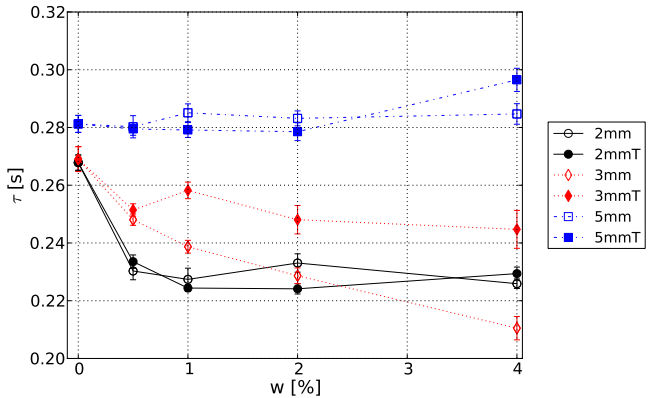}
\caption{\label{tau} (Color online) Dependence of the time span $\tau$  on liquid content, for different particle size and surface tension values.}
\end{figure}

As regards the time span of the collapse, $\tau$, which results are collected in Figure \ref{tau}, there is again a clear distinction between small and coarse particles. For 5 mm particles $\tau$ seems not to be affected by the presence of liquid nor by its surface tension, while smaller particles show a large dependence on water content and, for the 3 mm case, on liquid surface tension. Analysis of the dynamics therefore suggests that some clear patterns exist, which are however complicated by the various phenomena involved (friction, capillarity, finite size effects, eventually lubrication of contacts, etc.). 

\section{Discussion}

Results presented in the previous section highlighted an important feature of wet granular media: While in dry conditions the collapse phenomenon seems to be independent of particle size, the addition of a wetting fluid breaks this invariance inducing a dependence on particle diameter and surface tension. Moreover, a dependence on liquid amount was found for $w<1\%$.

The emergence of a dependence on particle size is obviously related to the effect of particle size on capillary forces. 
Dimensional analysis can help in understanding and comparing the importance of the phenomena involved in the experiments. The simplest dimensional number can be constructed by comparing capillary and body forces. This is the Bond number, which is defined as:
\be
\mbox{Bo}=\frac{\rho g d_p^2}{\gamma}
\ee

In the experimental campaign, two parameters forming Bond number were varied independently: particle diameter $d_p$ and liquid surface tension $\gamma$.
If Bond number was sufficient for scaling, we would find that lowering the surface tension by a factor of 4 would be equal to doubling the particle diameter. That is, 2 mm particles wetted with the low surface tension liquid should behave nearly the same as 5 mm particles for the larger $\gamma$: that is not too far from what was found for $L^*$, $H^*$ and $\theta_{top}$.  However, as discussed in previous sections, particularly for $w<1\%$ a dependence of the system behavior on water amount was clearly highlighted. Therefore Bond number alone cannot explain properly the experimental results, since it does not contain information on the liquid content. Bond number simply compares body and capillary forces on two particles in contact (for which the capillary force does not depend on bridge volume). Detailed empirical or analytical models of particle-particle interactions could be adopted to introduce the effect of $w$ on the strength of a capillary bridge, with the purpose of constructing a more refined dimensionless number.
For example \citet{lambert08} provide a theoretical formulation of the capillary force as function of the capillary volume and gap based on the minimum energy approach. Other empirical and semi-empirical relations can be found in \citet{willett00} and \citet{richefeu08}.  For a fixed liquid content and assuming a homogeneous distribution a mean capillary volume can be obtained.

However since runout dynamics is rather complex (multiple bridges exist for each particle, not all the contacts in principle support a bridge, etc.), simplifying assumptions should be adopted in order to adapt the single contact models to real experimental conditions. In particular, assumptions on the coordination number, on the average distance between particles and on the distribution of bridge volumes should be done. To avoid the complications related to such a detailed approach, and with the aim of finding scaling laws as simple as possible, we prefer to look for a combination of dimensionless numbers $\mbox{Bo}^mw^n$ as a scaling parameter.

\begin{figure}[ht]
\centering
\includegraphics[width=\columnwidth]{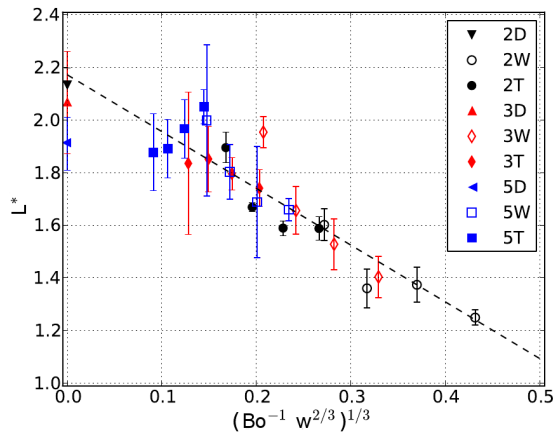}
\caption{\label{bo1}(Color online) Dependence of the dimensionless runout length $L^*$  on the dimensionless number $\mbox{Bo}^{-1}w^{2/3}$. The number in the legend denotes particle diameter, while the letter denotes: D the dry case, W the case with distilled water, T the case with water and a surfactant.}
\end{figure}

In particular, focusing on the runout length, as reported in Figure \ref{bo1}, it was found that raw data can be progressively reduced to a master curve when plotted against  $\mbox{Bo}^{-1}w^{2/3}$. The master curve expression is $L^*=2.17 \left[  1- \left(\mbox{Bo}^{-1}w^{2/3}\right)^{1/3} \right]$. 5 mm spheres wetted with the low liquid surface tension fall on the same curve but do not scale correctly on water content, supporting the hypothesis that some additional effect has to be considered (for example, lubrication). A nice result is that also data from dry tests do not fall too far from the curve. 

\begin{figure}[ht]
\centering
\includegraphics[width=\columnwidth]{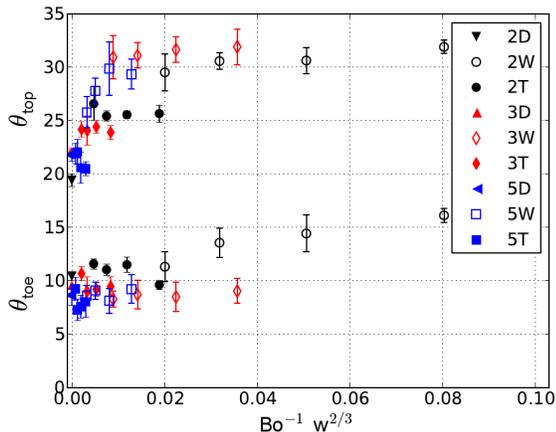}
\caption{\label{bo2} (Color online) Dependence of the top and toe angles on the dimensionless number $\mbox{Bo}^{-1}w^{2/3}$. The number in the legend denotes particle diameter, while the letter denotes: D the dry case, W the case with distilled water, T the case with water and a surfactant.}
\end{figure}

As reported in Figure \ref{bo2}, also the top and toe angles are described by the scaling parameter $\mbox{Bo}^{-1}w^{2/3}$. In particular, the scaling resumes quite well how these angles depend on particle diameter, surface tension and liquid amount. The 2/3 exponent on $w$ is clearly related to the ``surface'' nature of the capillary interactions, which suggests that the behavior will scale more likely with the wet area of contact,  which is proportional to $w^{2/3}$, than with the liquid bridge volume, which is proportional to $w$.

Another parameter which it would be useful to characterize with dimensional analysis is the maximum velocity of the collapsing event, $V_0$. 
Various permutations of variables forming dimensionless numbers with $V_0$ were tested. Some examples are the particle scale Froude number $\mbox{Fr}_{d_p}= \frac{V_0}{\sqrt{g d_p}}$, the flow scale Froude number, e.g. $\mbox{Fr}_H= \frac{V_0}{\sqrt{g H_0}}$, or more complex numbers taking into account particle-geometry interactions, like, e.g. $\mbox{Fr}^*= \frac{V_0}{\sqrt{g H_0}}\left(\frac{d_p}{W_0}\right)^m$. Unfortunately a satisfying scaling was not found, and more data should be collected, particularly for different aspect ratios of the initial column. As already suggested, the difficulty in finding a general behavior may come from finite size effects. To overcome this issues, axisymmetrical collapse experiments should be performed, as it was done for the dry column collapse \cite{lube04,lajeunesse04}. 

\section{Conclusions}

The focus of this work was the experimental characterization of the collapse of wet granular columns in the pendular state. The final deposit shape and the runout dynamics were studied for samples of glass beads, varying particle diameter, liquid surface tension and liquid amount. Data generally showed that the runout length decreases with increasing water amount  (reaching a plateaux for $w> 1\%$) and increases with increasing Bond number, while the top and toe angles, and the final deposit height increase with increasing water amount and decrease with decreasing Bond number.
Some differences were always observed for the largest Bond number case, where probably lubrication effects take place. Probable size effects were highlighted regarding the runout maximum velocity.

Dimensional analysis allowed to discuss possible scalings for the runout length and the top and toe angles: A satisfying scaling was found, based on the dimensionless group $\mbox{Bo}^{-1}w^{2/3}$. 

This work aims to be a first step in the understanding of the effect of capillary bridges on the mechanics of dense granular flows, and particularly on the collapse dynamics. The data presented in this work can be used as a reference for further studies, in order to develop refined scaling laws; on the other hand, it can be useful to compare collapse data to numerical simulations of wet granular collapse in order to validate micromechanical models of capillary bridge dynamics.

\begin{acknowledgements}
The authors gratefully acknowledge the Surfaces and Interfaces Physics Laboratory Group of the University of Padova and particularly Giorgio Delfitto, Davide Ferraro and Tamara Toth for the contact angle measurements, and for the helpful discussions on surface wettability.
\end{acknowledgements}


\begin{thebibliography}{19}
\expandafter\ifx\csname natexlab\endcsname\relax\def\natexlab#1{#1}\fi
\expandafter\ifx\csname bibnamefont\endcsname\relax
  \def\bibnamefont#1{#1}\fi
\expandafter\ifx\csname bibfnamefont\endcsname\relax
  \def\bibfnamefont#1{#1}\fi
\expandafter\ifx\csname citenamefont\endcsname\relax
  \def\citenamefont#1{#1}\fi
\expandafter\ifx\csname url\endcsname\relax
  \def\url#1{\texttt{#1}}\fi
\expandafter\ifx\csname urlprefix\endcsname\relax\def\urlprefix{URL }\fi
\providecommand{\bibinfo}[2]{#2}
\providecommand{\eprint}[2][]{\url{#2}}

\bibitem[{\citenamefont{Lube et~al.}(2004)\citenamefont{Lube, Huppert, Sparks,
  and Hallworth}}]{lube04}
\bibinfo{author}{\bibfnamefont{G.}~\bibnamefont{Lube}},
  \bibinfo{author}{\bibfnamefont{H.~E.} \bibnamefont{Huppert}},
  \bibinfo{author}{\bibnamefont{Sparks}}, \bibnamefont{and}
  \bibinfo{author}{\bibfnamefont{M.~A.} \bibnamefont{Hallworth}},
  \bibinfo{journal}{Journal of Fluid Mechanics} \textbf{\bibinfo{volume}{508}},
  \bibinfo{pages}{175} (\bibinfo{year}{2004}).

\bibitem[{\citenamefont{Lajeunesse et~al.}(2004)\citenamefont{Lajeunesse,
  Mangeney-Castelnau, and A.}}]{lajeunesse04}
\bibinfo{author}{\bibfnamefont{E.}~\bibnamefont{Lajeunesse}},
  \bibinfo{author}{\bibnamefont{Mangeney-Castelnau}}, \bibnamefont{and}
  \bibinfo{author}{\bibfnamefont{J.}~\bibnamefont{A.}, \bibfnamefont{Vilotte}},
  \bibinfo{journal}{Physics of Fluids} \textbf{\bibinfo{volume}{16}},
  \bibinfo{pages}{2371} (\bibinfo{year}{2004}).

\bibitem[{\citenamefont{Lube et~al.}(2005)\citenamefont{Lube, Huppert, Sparks,
  and Freundt}}]{lube05}
\bibinfo{author}{\bibfnamefont{G.}~\bibnamefont{Lube}},
  \bibinfo{author}{\bibfnamefont{H.}~\bibnamefont{Huppert}},
  \bibinfo{author}{\bibfnamefont{R.}~\bibnamefont{Sparks}}, \bibnamefont{and}
  \bibinfo{author}{\bibfnamefont{A.}~\bibnamefont{Freundt}},
  \bibinfo{journal}{Physical Review E - Statistical, Nonlinear, and Soft Matter
  Physics} \textbf{\bibinfo{volume}{72}}, \bibinfo{pages}{1}
  (\bibinfo{year}{2005}).

\bibitem[{\citenamefont{Siavoshi and Kudrolli}(2005)}]{siavoshi05b}
\bibinfo{author}{\bibfnamefont{S.}~\bibnamefont{Siavoshi}} \bibnamefont{and}
  \bibinfo{author}{\bibfnamefont{A.}~\bibnamefont{Kudrolli}},
  \bibinfo{journal}{Physical Review E - Statistical, Nonlinear, and Soft Matter
  Physics} \textbf{\bibinfo{volume}{71}}, \bibinfo{pages}{1}
  (\bibinfo{year}{2005}).

\bibitem[{\citenamefont{Lajeunesse et~al.}(2005)\citenamefont{Lajeunesse,
  Monnier, and Homsy}}]{lajeunesse05}
\bibinfo{author}{\bibfnamefont{E.}~\bibnamefont{Lajeunesse}},
  \bibinfo{author}{\bibfnamefont{J.}~\bibnamefont{Monnier}}, \bibnamefont{and}
  \bibinfo{author}{\bibfnamefont{G.}~\bibnamefont{Homsy}},
  \bibinfo{journal}{Physics of Fluids} \textbf{\bibinfo{volume}{17}}
  (\bibinfo{year}{2005}).

\bibitem[{\citenamefont{Balmforth and Kerswell}(2005)}]{balmforth05}
\bibinfo{author}{\bibfnamefont{N.}~\bibnamefont{Balmforth}} \bibnamefont{and}
  \bibinfo{author}{\bibfnamefont{R.}~\bibnamefont{Kerswell}},
  \bibinfo{journal}{Journal of Fluid Mechanics} \textbf{\bibinfo{volume}{538}},
  \bibinfo{pages}{399} (\bibinfo{year}{2005}).

\bibitem[{\citenamefont{Zenit}(2005)}]{zenit05}
\bibinfo{author}{\bibfnamefont{R.}~\bibnamefont{Zenit}},
  \bibinfo{journal}{Physics of Fluids} \textbf{\bibinfo{volume}{17}},
  \bibinfo{pages}{031703} (\bibinfo{year}{2005}).

\bibitem[{\citenamefont{STARON and HINCH}(2005)}]{staron05b}
\bibinfo{author}{\bibfnamefont{L.}~\bibnamefont{Staron}} \bibnamefont{and}
  \bibinfo{author}{\bibfnamefont{E.~J.} \bibnamefont{Hinch}},
  \bibinfo{journal}{Journal of Fluid Mechanics} \textbf{\bibinfo{volume}{545}},
  \bibinfo{pages}{1} (\bibinfo{year}{2005}).

\bibitem[{\citenamefont{Staron and Hinch}(2007)}]{staron07}
\bibinfo{author}{\bibfnamefont{L.}~\bibnamefont{Staron}} \bibnamefont{and}
  \bibinfo{author}{\bibfnamefont{E.}~\bibnamefont{Hinch}},
  \bibinfo{journal}{Granular Matter} \textbf{\bibinfo{volume}{9}},
  \bibinfo{pages}{205} (\bibinfo{year}{2007}).

\bibitem[{\citenamefont{Mangeney-Castelnau
  et~al.}(2005)\citenamefont{Mangeney-Castelnau, Bouchut, Vilotte, Lajeunesse,
  Aubertin, and Pirulli}}]{mangeney05}
\bibinfo{author}{\bibfnamefont{A.}~\bibnamefont{Mangeney-Castelnau}},
  \bibinfo{author}{\bibfnamefont{F.}~\bibnamefont{Bouchut}},
  \bibinfo{author}{\bibfnamefont{J.}~\bibnamefont{Vilotte}},
  \bibinfo{author}{\bibfnamefont{E.}~\bibnamefont{Lajeunesse}},
  \bibinfo{author}{\bibfnamefont{A.}~\bibnamefont{Aubertin}}, \bibnamefont{and}
  \bibinfo{author}{\bibfnamefont{M.}~\bibnamefont{Pirulli}},
  \bibinfo{journal}{Journal of Geophysical Research B: Solid Earth}
  \textbf{\bibinfo{volume}{110}}, \bibinfo{pages}{1} (\bibinfo{year}{2005}).

\bibitem[{\citenamefont{Kerswell}(2005)}]{kerswell05}
\bibinfo{author}{\bibfnamefont{R.}~\bibnamefont{Kerswell}},
  \bibinfo{journal}{Physics of Fluids} \textbf{\bibinfo{volume}{17}},
  \bibinfo{pages}{1} (\bibinfo{year}{2005}).

\bibitem[{\citenamefont{Larrieu et~al.}(2006)\citenamefont{Larrieu, Staron, and
  Hinch}}]{larrieu06}
\bibinfo{author}{\bibfnamefont{E.}~\bibnamefont{Larrieu}},
  \bibinfo{author}{\bibfnamefont{L.}~\bibnamefont{Staron}}, \bibnamefont{and}
  \bibinfo{author}{\bibfnamefont{E.}~\bibnamefont{Hinch}},
  \bibinfo{journal}{Journal of Fluid Mechanics} \textbf{\bibinfo{volume}{554}},
  \bibinfo{pages}{259} (\bibinfo{year}{2006}).

\bibitem[{\citenamefont{Doyle et~al.}(2007)\citenamefont{Doyle, Huppert, Lube,
  Mader, and Sparks}}]{doyle07}
\bibinfo{author}{\bibfnamefont{E.}~\bibnamefont{Doyle}},
  \bibinfo{author}{\bibfnamefont{H.}~\bibnamefont{Huppert}},
  \bibinfo{author}{\bibfnamefont{G.}~\bibnamefont{Lube}},
  \bibinfo{author}{\bibfnamefont{H.}~\bibnamefont{Mader}}, \bibnamefont{and}
  \bibinfo{author}{\bibfnamefont{R.}~\bibnamefont{Sparks}},
  \bibinfo{journal}{Physics of Fluids} \textbf{\bibinfo{volume}{19}}
  (\bibinfo{year}{2007}).

\bibitem[{\citenamefont{Meriaux and Triantafillou}(2008)}]{meriaux08}
\bibinfo{author}{\bibfnamefont{C.}~\bibnamefont{Meriaux}} \bibnamefont{and}
  \bibinfo{author}{\bibfnamefont{T.}~\bibnamefont{Triantafillou}},
  \bibinfo{journal}{Physics of Fluids} \textbf{\bibinfo{volume}{20}}
  (\bibinfo{year}{2008}).

\bibitem[{\citenamefont{Hornbaker et~al.}(1997)\citenamefont{Hornbaker, Albert,
  Albert, Barabasi, and Schiffer}}]{hornbaker97}
\bibinfo{author}{\bibfnamefont{D.~J.} \bibnamefont{Hornbaker}},
  \bibinfo{author}{\bibfnamefont{R.}~\bibnamefont{Albert}},
  \bibinfo{author}{\bibfnamefont{I.}~\bibnamefont{Albert}},
  \bibinfo{author}{\bibfnamefont{A.-L.} \bibnamefont{Barabasi}},
  \bibnamefont{and} \bibinfo{author}{\bibfnamefont{P.}~\bibnamefont{Schiffer}},
  \bibinfo{journal}{Nature} \textbf{\bibinfo{volume}{387}},
  \bibinfo{pages}{765} (\bibinfo{year}{1997}).

\bibitem[{\citenamefont{Newitt and Conway-Jones}(1958)}]{newitt58}
\bibinfo{author}{\bibfnamefont{D.~M.} \bibnamefont{Newitt}} \bibnamefont{and}
  \bibinfo{author}{\bibfnamefont{J.~M.} \bibnamefont{Conway-Jones}},
  \bibinfo{journal}{Trans. Inst. Chem. Eng.} \textbf{\bibinfo{volume}{36}},
  \bibinfo{pages}{422} (\bibinfo{year}{1958}).
  
\bibitem[{\citenamefont{Richefeu et~al.}(2008)\citenamefont{Richefeu,
  El~Youssoufi, Peyroux, and Radja{\"\i}}}]{richefeu08}
\bibinfo{author}{\bibfnamefont{V.}~\bibnamefont{Richefeu}},
  \bibinfo{author}{\bibfnamefont{M.~S.} \bibnamefont{El~Youssoufi}},
  \bibinfo{author}{\bibfnamefont{R.}~\bibnamefont{Peyroux}}, \bibnamefont{and}
  \bibinfo{author}{\bibfnamefont{F.}~\bibnamefont{Radja{\"\i}}},
  \bibinfo{journal}{International Journal for Numerical and Analytical Methods
  in Geomechanics} \textbf{\bibinfo{volume}{32}}, \bibinfo{pages}{1365}
  (\bibinfo{year}{2008}), ISSN \bibinfo{issn}{1096-9853}.

\bibitem[{\citenamefont{Xu et~al.}(2007)}]{xu07}
\bibinfo{author}{\bibfnamefont{Q.}~\bibnamefont{Xu}},
  \bibinfo{author}{\bibfnamefont{A.~V.}~\bibnamefont{Orpe}} \bibnamefont{and}
  \bibinfo{author}{\bibfnamefont{A.}~\bibnamefont{Kudrolli}},
  \bibinfo{journal}{Physical Review E - Statistical, Nonlinear, and Soft Matter
  Physics} \textbf{\bibinfo{volume}{76}}, \bibinfo{pages}{031302}
  (\bibinfo{year}{2007}).
  
  
\bibitem[{\citenamefont{Lambert et~al.}(2008)\citenamefont{Lambert, Chau,
  Delchambre, and Regnier}}]{lambert08}
\bibinfo{author}{\bibfnamefont{P.}~\bibnamefont{Lambert}},
  \bibinfo{author}{\bibfnamefont{A.}~\bibnamefont{Chau}},
  \bibinfo{author}{\bibfnamefont{A.}~\bibnamefont{Delchambre}},
  \bibnamefont{and} \bibinfo{author}{\bibfnamefont{S.}~\bibnamefont{Regnier}},
  \bibinfo{journal}{Langmuir} \textbf{\bibinfo{volume}{24}},
  \bibinfo{pages}{3157} (\bibinfo{year}{2008}).
  
\bibitem[{\citenamefont{Willett et~al.}(2000)\citenamefont{Willett, Adams,
  Johnson, and Seville}}]{willett00}
\bibinfo{author}{\bibfnamefont{C.~D.} \bibnamefont{Willett}},
  \bibinfo{author}{\bibfnamefont{M.~J.} \bibnamefont{Adams}},
  \bibinfo{author}{\bibfnamefont{S.~A.} \bibnamefont{Johnson}},
  \bibnamefont{and} \bibinfo{author}{\bibfnamefont{J.~P.~K.}
  \bibnamefont{Seville}}, \bibinfo{journal}{Langmuir}
  \textbf{\bibinfo{volume}{16}}, \bibinfo{pages}{9396} (\bibinfo{year}{2000}).
  


\end{thebibliography}

\end{document}